\documentclass[aps,prb,twocolumn,showpacs,%
preprintnumbers,amssymb]{revtex4-1}
\usepackage{graphicx}
\usepackage[colorlinks=true,linkcolor=blue,%
pagecolor=blue,citecolor=blue,%
urlcolor=blue,anchorcolor=black,%
bookmarksnumbered=true,%
bookmarksopen=true]{hyperref}
\usepackage{dcolumn}
\usepackage{bm}
\renewcommand{\frac}[2]{\displaystyle{#1 \over #2}}
\begin{document}
\preprint{\underline{\href{%
http://dx.doi.org/10.1063/1.4978561}%
{Physics of Plasmas, 2017, vol.\ 24, no.\ 3, 
pp.\ 033709-1--033709-7.}}}
\title{Driving force for a nonequilibrium phase 
transition in three-dimensional complex 
plasmas}
\author{D.~I.~Zhukhovitskii} \email{dmr@ihed.ras.ru}
\homepage{http://oivtran.ru/dmr/}
\affiliation{Joint Institute of High Temperatures, Russian 
Academy of Sciences, Izhorskaya 13, Bd.~2, 125412 
Moscow, Russia}
\date{\today}
\begin{abstract}
An example of the non-equilibrium phase transition is 
the formation of lanes when one kind of particles is 
driven against the other. According to experimental 
observation, lane formation in binary complex plasmas 
occurs when the smaller particles are driven through the 
stationary dust cloud of the larger particles. We calculate 
the driving force acting on a probe particle that finds 
itself in a quiescent cloud of particles in complex plasma 
of the low-pressure radio frequency discharge under 
microgravity conditions. It is shown that the nonzero 
driving force is a result of the dependence of the ion 
mean free path on the particle number density. If this 
effect is properly included in the model of similar 
complex plasmas then one arrives at the driving force 
that changes its sign at the point where the probe and the 
dust particles have equal radii. If the probe is smaller 
than the dust particle then the driving force is directed 
toward the discharge center and vice versa, in 
accordance with experiment. Obtained results can serve 
as the ansatz for future investigation of the lane 
formation in complex plasmas.
\end{abstract}
\maketitle
\section{\label{s1}INTRODUCTION}

Complex plasmas are low-temperature plasmas including 
dust particles, typically in the micrometer range. These are 
dusty plasmas, which are specially prepared to study 
fundamental processes in the strong coupling regime on the 
most fundamental level when the observation of individual 
motion of particles and their interactions is possible. Under 
microgravity conditions realized either in parabolic 
flights\cite{10,11,12,13,14} or onboard the International 
Space Station (ISS),\cite{10,15,16,17,18,019,19} such 
relatively weak forces as the ion drag force and the 
interparticle interactions become important and often 
determine the motion and structure formation in complex 
plasma. Since the mobility of electrons is much greater than 
that of ions, particles acquire a significant negative electric 
charge. This leads to formation of a strongly coupled 
plasma,\cite{1,2,3,4,5,6,8,9} in which large volumes of 
almost homogeneous three-dimensional complex plasma 
can be observed.

This strongly coupled open system can exhibit a number of 
nonequilibrium phase transitions. Among them, one of the 
most vivid effects is the lane formation when two kinds of 
particles are driven against each other. If the driving forces 
are strong enough like particles form ``stream lines'' and 
move collectively in lanes, which show an anisotropic 
structural order accompanied by a considerable 
enhancement of the particle mobility. Examples of this 
phenomenon include driven bi-layer systems and 
two-species lattice gases,\cite{72} granular 
mixtures,\cite{74} molecular ions,\cite{83} highly 
populated pedestrian zones,\cite{73} and driven colloidal 
mixtures.\cite{75,76,77,78,79,80,81,82} Lane formation in 
three-dimensional complex plasma of a dust cloud in the 
radio frequency low-pressure gas discharge was observed in 
Refs.~\onlinecite{68,63}.

Lane formation in complex plasmas is a special issue for 
two reasons. First, only one species of the dust particles is 
driven while the second one is quiescent. Note that in this 
case, the only difference between the two species is the 
particle radius. Obviously, this peculiarity is not of 
principal nature. Second, the driving force is not an external 
one but it is a sum of the internal forces acting on the 
particle in plasma. In most cases, these forces are the 
oppositely directed electric and ion drag forces, which 
cannot be directly controlled. Moreover, while, e.g., the 
driving force in colloidal mixtures is treated as a 
preassigned parameter, it is not known for complex 
plasmas. Thus, calculation of the driving force becomes a 
separate problem, which is addressed in this work.

The force acting on an isolated dust particle is 
well-known.\cite{1} However, this approach is 
inappropriate for the strongly coupled dust cloud where the 
interparticle interactions are quite essential. In the recent 
studies,\cite{22,64} the electric and ion drag forces are 
calculated assuming that the Coulomb potentials of 
neighboring particles overlap. This provides an 
interpretation for the particle number density in a dust cloud 
and determines a relationship between the number densities 
of all plasma charge carriers and the particle charge. We 
will term the force acting on an individual probe particle of 
the radius $a_p$ that finds itself in a dust cloud of particles 
of the radius $a$
 the driving force. Surprisingly, calculation of this force 
based on the model of similar complex plasmas 
(SP)\cite{64} results in the force vanishing.

In this work, we will show that the {\it nonzero\/} driving 
force is a net result of the fact that the ion mean free path 
with respect to the collisions against neutrals can be 
comparable with the interparticle distance, which was not 
taken into account in Ref.~\onlinecite{64}. If we calculate 
the total mean free path with due regard for the ion 
scattering on the dust particles then the ionization equation 
of state (IEOS) is modified so that the driving force does 
not vanish. In this way, we derive the modified model of 
similar complex plasmas (MSP). The calculation using 
MSP shows that if $a_p > a$, the larger {\it subsonic\/} 
particle that we term the probe moves through the dust 
cloud {\it along\/} the ion flux toward the outer boundary 
of a dust cloud, in correspondence with the experiments. 
For the larger supersonic particle, the estimations show that 
due to the absence of a spherical cavity around the moving 
particle the ion drag force is much weaker than in the case 
of a subsonic particle. Hence, the driving force is not much 
different from the electric force and the probe velocity is 
directed against the ion drag force toward the discharge 
center. The developed approach can be used as an ansatz 
for development of the theory of lane formation in complex 
plasmas.

The paper is organized as follows. In Sec.~\ref{s2}, the 
IEOS for the stationary dust cloud obtained in our previous 
studies is generalized to include the effect of finite ion 
mean free path affected by the collisions against dust 
particles. In Sec.~\ref{s3}, we explore the modified IEOS 
and derive the pressure of the dust particle subsystem 
necessary for the calculation of driving force and check the 
effect of IEOS modification on the velocity of dust acoustic 
waves. The driving force is calculated in Sec.~\ref{s4}, and 
the resulting larger particle velocity is estimated and 
compared with experiments in Sec.~\ref{s5}. The results of 
this study are summarized in Sec.~\ref{s6}, and the 
parameters of a cavity behind the supersonic probe are 
estimated in the Appendix~\ref{sA}.

\section{\label{s2}MODIFIED IONIZATION 
EQUATION OF STATE}

Consider a stationary dust cloud in the low-pressure gas 
discharge. Here and in Sec.~\ref{s3}, we imply that the 
particles forming a cloud have the same radius. Under 
microgravity conditions, a dust particle is subject to three 
basic forces, namely, the electric driving force, the ion drag 
force arising from scattering of the streaming ions on dust 
particles, and the neutral drag force (friction force) due to 
collisions of the atoms against the moving particles. For a 
stationary cloud, the latter force vanishes. Note that in a 
strongly coupled system, the correlation energy originating 
from particle ordering results in the difference between the 
volume-averaged electric field and the electric field at the 
point of particle location. This effect can be included if we 
introduce the dust pressure. The effect of this pressure on 
the force balance equation in a stationary plasma is in most 
cases negligible; however, it is responsible for propagation 
of a perturbation.\cite{64} The electric field driving force 
${\bf{f}}_e$ and the ion drag force ${\bf{f}}_{id}$ acting 
on unit volume can be written in the form
\begin{equation}
{\bf{f}}_e = - Zen_d {\bf{E}} = - \frac{{aT_e }}{e}\Phi 
n_d {\bf{E}}, \label{e1}
\end{equation}
where $Z$
 is the dust particle charge in units of the electron charge, 
$e$
 is the elementary electric charge, $n_d$ is the particle 
number density, $a$
 is the particle radius, $T_e$ is the electron temperature, 
$\Phi = Ze^2 /aT_e$ is the dimensionless potential of a dust 
particle, ${\bf{E}} = (T_e /e)\bm{\nabla}\ln n_e$ is the 
electric field strength, $n_e$ is the electron number density, 
and
\begin{equation}
{\bf{f}}_{id} = \frac{3}{8}\left( {\frac{{4\pi }}{3}} 
\right)^{1/3} \beta ^{1/3} n_d^{1/3} n_i \lambda 
e{\bf{E}}, \label{e2}
\end{equation}
where $\lambda$ is the ion mean free path, $n_i$ is the ion 
number density, and $\beta$ is a dimensionless free 
parameter. Equation (\ref{e2}) implies that the cross 
section of momentum transfer from the ions to dust 
particles is $\sigma _{\mathrm{eff}} = (\pi /2)\beta ^{1/3} 
r_d^2 $, where $r_d = (3/4\pi n_d )^{1/3}$ is the dust 
particle Wigner--Seitz cell radius. The coefficient $\beta$ 
depends on $\Phi $, $r_d $, and the ion temperature $T_i$ 
(in Refs.~\onlinecite{22,64}, $\sigma _{\mathrm{eff}}$ 
was estimated using the particle charge screening length in 
the Wigner--Seitz cell $ \simeq 0.45r_d$ that corresponds 
to $\beta = 1$).

The force balance equation reads ${\bf{f}}_e + 
{\bf{f}}_{id} = 0$
 or
\begin{equation}
\frac{\pi }{2}\beta ^{1/3} r_d^2 n_i \lambda = \frac{{aT_e 
}}{{e^2 }}\Phi . \label{e3}
\end{equation}
The ion mean free path $\lambda$ appearing in 
Eq.~(\ref{e3}) is defined by the collisions both with the gas 
atoms and with the dust particles. In contrast to 
Refs.~\onlinecite{22,64}, we take this into account and 
calculate $\lambda$ as
\begin{equation}
\lambda ^{ - 1} = \lambda _a^{ - 1} + \sigma 
_{\mathrm{eff}} n_d = \lambda _a^{ - 1} \left( {1 + 
\frac{3}{8}n_d^{*1/3} } \right), \label{e4}
\end{equation}
where $\lambda _a$ is the ion mean free path with respect 
to the collisions against gas atoms (in a gas discharge 
without particles) and $n_d^* = (4\pi /3)\beta n_d \lambda 
_a^3$ is the dimensionless particle number density.

The combination of Eqs.~(\ref{e3}) and (\ref{e4}) with the 
equation for particle potential that follows from the orbital 
motion limited (OML) approximation\cite{54,55} at $T_e 
/T_i \gg 1$
\begin{equation}
\theta \Phi e^\Phi  = \frac{{n_e }}{{n_i }}, \label{e5}
\end{equation}
where $\theta = \sqrt {T_e m_e /T_i m_i } $, $m_e$ and 
$m_i$ is the electron and ion mass, respectively, $n_e$ is 
the electron number density and the local quasineutrality 
condition
\begin{equation}
n_i = \frac{{aT_e }}{{e^2 }}\Phi n_d + n_e , \label{e6}
\end{equation}
yield the modified IEOS
\begin{equation}
\theta \Phi e^\Phi  + \frac{3}{8}\left( {\frac{{\pi \tilde 
n_i^* }}{{2\Phi }}} \right)^{1/2} = 1, \label{e7}
\end{equation}
where
\begin{equation}
\tilde n_i^{*1/3} = \frac{{n_i^{*1/3} }}{{1 + 
\frac{3}{8}\left( {\frac{{4\pi }}{3}\frac{{\gamma n_i^* 
}}{\Phi }} \right)^{1/3} }}, \label{e8}
\end{equation}
$n_i^* = \beta e^2 \lambda _a^3 n_i /aT_e$ is the 
dimensionless ion number density, and $\gamma (\Phi ) = 1 
- \theta \Phi e^\Phi $. If we replace $\tilde n_i^*$ by 
$n_i^*$ in Eq.~(\ref{e7}) it would coincide with Eq.~(12) 
in Ref.~\onlinecite{64}, i.e., the initial and modified IEOS 
have the same form. From Eqs.~(\ref{e7}) and (\ref{e8}), 
the ion number density can be represented as an explicit 
function of $\Phi$\begin{equation}
n_i^* = \frac{{128}}{{9\pi }}\frac{{\Phi \gamma ^2 
}}{{(1 - \gamma )^3 }}. \label{e0080}
\end{equation}

Given $\Phi$ or $n_i^* $, one can calculate the particle 
number density
\begin{equation}
n_d^* = \frac{{4\pi }}{3}\frac{{\gamma n_i^* }}{\Phi } 
\label{e9}
\end{equation}
and the dimensionless electron number density $n_e^* = 
\beta e^2 \lambda _a^3 n_e /aT_e $,
\begin{equation}
n_e^* = n_i^* \theta \Phi e^\Phi  = \frac{2}{\pi }\Phi 
n_d^{*2/3} . \label{e001}
\end{equation}
\begin{figure}
\includegraphics[width=10cm]{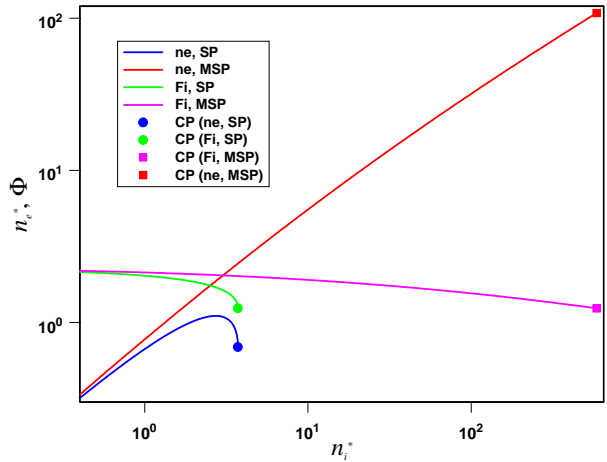}
\caption{\label{f1}Dimensionless electron number density 
$n_e^*$ from SP (blue line) and MSP (red line) and the 
particle potential $\Phi$ from SP (green line) and MSP 
(magenta line) as a function of the dimensionless ion 
number density $n_i^*$ for $\theta = 
{\mbox{0}}{\mbox{.0431}}$. Solid dots indicate positions 
of the critical points in SP (circles) and MSP (squares).}
\end{figure}
\begin{figure}
\includegraphics[width=10cm]{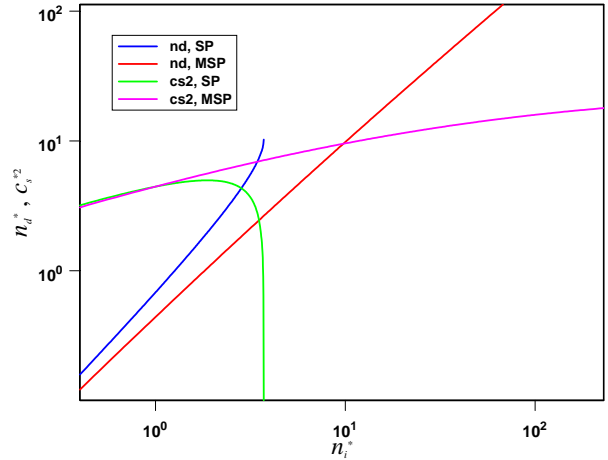}
\caption{\label{f2}Dimensionless particle number density 
$n_d^*$ from SP (blue line) and MSP (red line) and the 
squared sound velocity $c_s^2$ from SP (green line) and 
MSP (magenta line) as a function of the dimensionless ion 
number density $n_i^* $, $\theta = 
{\mbox{0}}{\mbox{.0431}}$.}
\end{figure}
Figures~\ref{f1} and \ref{f2} compare the plasma 
parameters calculated using Eq.~(12) in 
Ref.~\onlinecite{64} (SP) and Eq.~(\ref{e0080}) (MSP). It 
is seen that at low number densities of the charged 
components, the calculation results almost coincide, while 
the maximum $n_i^*$ that satisfies the modified IEOS 
(\ref{e0080}) is significantly greater. The same is true for 
$n_e^*$ and $n_d^* $. A remarkable difference is 
observed between the critical points in SP and MSP 
(Fig.~\ref{f1}). By definition, at the critical point, $n_i^*$ 
reaches maximum and the cloud is stable, $dp^* /dn_d^* > 
0$, where $p^*$ is the dimensionless pressure (see 
Sec.~\ref{s3}). For SP, $dn_i^* /dn_e^* = dn_i^* /d\Phi = 
0$
 at the critical point while for MSP, these derivatives do not 
vanish. The critical $n_i^*$ is much greater for MSP than 
for SP. Obviously, such number densities are not realistic 
because the rate of ion-electron recombination on the 
particle surface would be too high to sustain the discharge. 
At the same time, in contrast to SP,\cite{22} MSP does 
account for high particle number densities in the vicinity of 
the void boundary. Recall that the void is the dust particle 
free region in the center of the discharge.\cite{1}

\section{\label{s3}DUST PARTICLE PRESSURE AND 
THE SOUND VELOCITY}

\begin{figure}
\includegraphics[width=10cm]{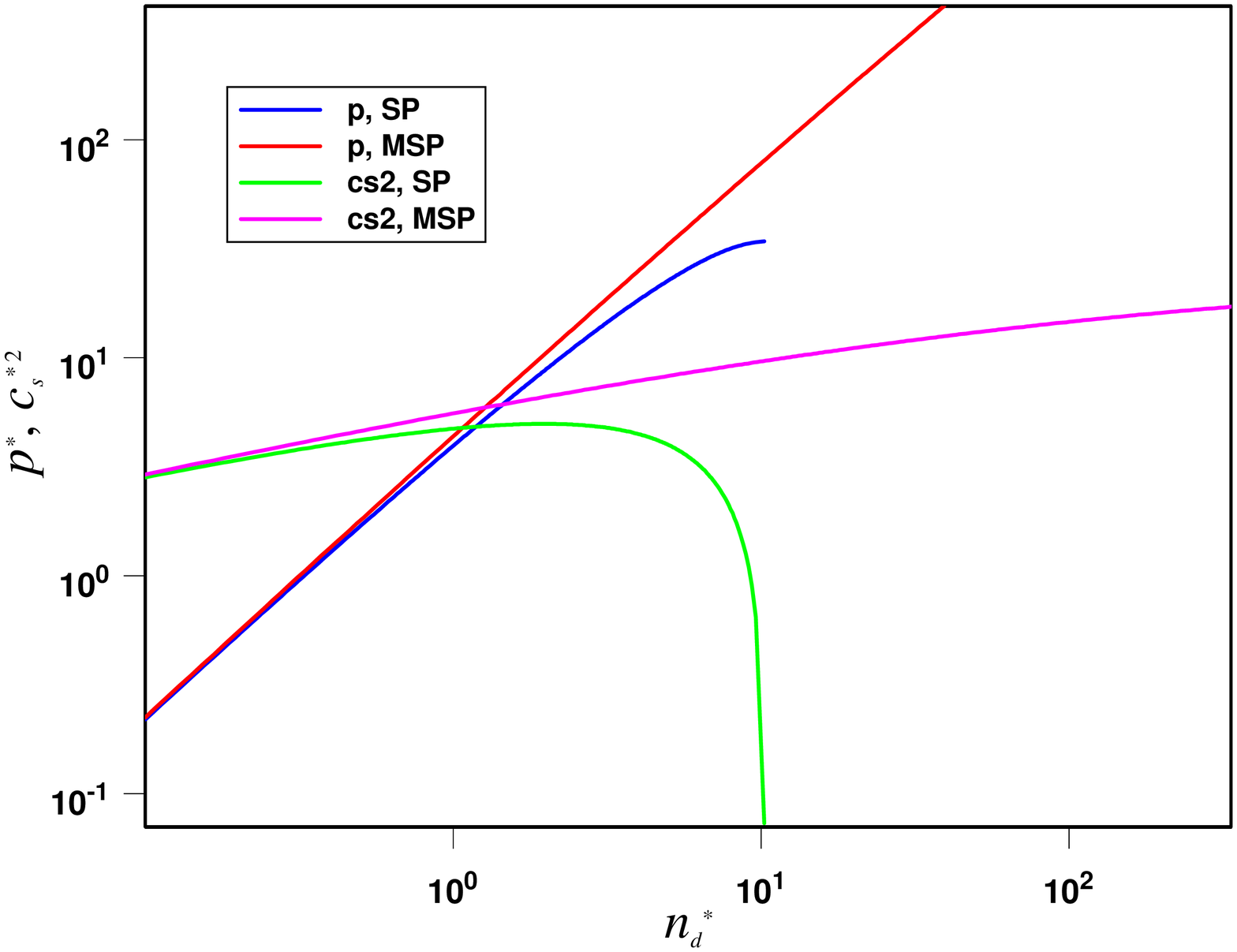}
\caption{\label{f3}Dimensionless dust particle pressure 
$p^*$ from SP (blue line) and MSP (red line) and the 
squared sound velocity $c_s^2$ from SP (green line) and 
MSP (magenta line) as a function of the dimensionless 
particle number density $n_d^* $, $\theta = 
{\mbox{0}}{\mbox{.0431}}$.}
\end{figure}

The dust particle pressure is defined as\cite{22}
\begin{equation}
p = \frac{{Z^2 e^2 }}{{8\pi r_d^4 }}. \label{e10}
\end{equation}
One can introduce the dimensionless pressure
\begin{equation}
p^* = \frac{{8\pi e^2 \lambda _a^4 \beta ^{4/3} }}{{a^2 
T_e^2 }}p = \Phi ^2 n_d^{*4/3} . \label{e11}
\end{equation}
The dependence $p^* (n_d^* )$
 is shown in Fig.~\ref{f3}. As is seen, for this quantity, the 
difference between SP and MSP is insignificant.

The IEOS for the variables $n_d^* ,\;\Phi$ follows 
straightforwardly from (\ref{e0080}) and (\ref{e9})
\begin{equation}
n_d^{*1/3} = \frac{{8\gamma }}{{3(1 - \gamma )}}. 
\label{e12}
\end{equation}
We introduce the dimensionless velocity of dust acoustic 
waves (sound velocity)
\begin{equation}
c_s^* = \frac{{c_s e(6M_d \lambda _a )^{1/2} \beta ^{1/6} 
}}{{aT_e }}, \label{e13}
\end{equation}
where $c_s$ is the sound velocity and $M_d$ is the mass of 
a dust particle. Since $c_s^2 = M_d^{ - 1} (dp/dn_d )$, we 
obtain from Eqs.~(\ref{e11})--(\ref{e13})
\begin{equation}
c_s^{*2} = \frac{{dp^* }}{{dn_d^* }} = \frac{4}{3}\Phi 
^2 n_d^{*1/3} \left[ {1 - \frac{\gamma }{{2(1 + \Phi )}}} 
\right]. \label{e14}
\end{equation}
Eq.~(\ref{e14}) differs from corresponding Eq.~(35) in 
Ref.~\onlinecite{64}. Although according to the modified 
IEOS (\ref{e0080}) the dust pressure is sensitive to $n_d^* 
$, the sound velocity is not much different from that 
obtained in Ref.~\onlinecite{64} (Figs.~\ref{f2} and 
\ref{f3}) and, therefore, from the experimental results. 
Likewise, it is almost independent on $n_i$ and $n_e $, i.e., 
on the position in the dust cloud. In contrast to 
Ref.~\onlinecite{64}, where the calculated $c_s$ vanishes 
at the critical point, such behavior extends to considerably 
higher $n_d^* $, so that this range covers the particle 
number densities typical for the void boundary. The fact 
that the number density variation has almost no effect on 
the sound velocity has been confirmed experimentally (see 
references in Ref.~\onlinecite{64}).

\section{\label{s4}CALCULATION OF THE DRIVING 
FORCE}

In this Section, we will treat a probe particle with the radius 
$a_p$ in the cloud of particles with the radius $a$. Such 
particles can appear in the discharge sporadically or they 
can be injected in plasma purposely. Our objective will be 
the calculation of the sum of the ion drag and electric force 
acting on this probe particle, which is the driving force. We 
will term the Wigner--Seitz cell of a probe ``cavity.'' If 
$R_p \gg a$, the cavity is the dust particle free region 
around a probe. The particles are displaced from the cavity 
due to the Coulomb repulsion between the probe and 
particles rather than to the ion drag force as in the case of 
the void. The radius of a cavity around the probe $R_p$ is 
obtained from the minimization of the work of its formation 
$(4\pi /3)pR_p^3 + Z_p^2 e^2 /2R_p $, where $Z_p$ is the 
charge of a probe particle in units of the electron 
charge,\cite{22}
\begin{equation}
R_p^2 = \frac{{Z_p e}}{{\sqrt {8\pi p} }} = \frac{{a_p 
}}{a}\frac{{\lambda _a^2 \beta ^{2/3} }}{{n_d^{*2/3} }}, 
\label{e15}
\end{equation}
and we used (\ref{e11}). It is remarkable that the cavity 
radius as a function of the dimensional particle number 
density
\begin{equation}
R_p = r_d \sqrt {\frac{{a_p }}{a}} \label{e003}
\end{equation}
is independent of $\beta $.

We will assume that at the boundary of a cavity around the 
probe particle, the electron and ion number densities 
coincide with $n_e$ and $n_i $, respectively, i.e., the 
potentials of dust and probe particles coincide. Hence, $\Phi 
= Z_p e^2 /a_p T_e $. If we define the direction of a 
coordinate axis $Y$
 apart from the void center as a positive direction then the 
electric force acting on the probe is $F_{ep} = a_p T_e \Phi 
E/e$, where $E = (T_e /e)(d\ln n_e /dy)$, $y$
 is the coordinate, and the ion drag force is $F_{ip} = - (\pi 
/2)\beta ^{1/3} R_p^2 n_i \lambda _p eE$, where
\begin{equation}
\lambda _p = \frac{{\lambda _a }}{{1 + 
\frac{3}{8}\frac{{\lambda _a }}{{R_p }}}} = 
\frac{{\lambda _a }}{{1 + \frac{\gamma }{{1 - \gamma 
}}\sqrt {\frac{a}{{a_p }}} }} \label{e004}
\end{equation}
is the ion mean free path in a cloud with the particle 
number density $(3/4\pi )R_p^{ - 3} $, and we used 
(\ref{e003}). With due regard for (\ref{e9}), the ratio of the 
absolute values $\kappa = - F_{ep} /F_{ip}$ is given by the 
following equation,
\begin{equation}
\kappa = 1 + \left( {\sqrt {\frac{a}{{a_p }}} - 1} 
\right)\gamma . \label{e16}
\end{equation}
It is noteworthy that $\kappa$ is independent of $\beta $. 
Thus, the sought driving force $F_{\mathrm{drv}} = 
F_{ep} + F_{ip}$ is
\begin{equation}
F_{\mathrm{drv}} = \frac{{a_p \Phi T_e^2 }}{{e^2 
}}\left( {1 - \frac{1}{\kappa }} \right)\frac{{d\ln n_e 
}}{{dy}}. \label{e17}
\end{equation}
If $\left| {a_p - a} \right|/a_p \ll 1$
 then (\ref{e17}) is reduced to
\begin{equation}
F_{\mathrm{drv}} = \frac{{\gamma \Phi T_e^2 (a - a_p 
)}}{{2e^2 }}\frac{{d\ln n_e }}{{dy}}. \label{e18}
\end{equation}
One can substitute $d\ln n_e /dy$
 in (\ref{e17}) and (\ref{e18}) by $d\ln n_d /dy$
 using (\ref{e001}). Since
\begin{equation}
\frac{{dn_e^* }}{{dn_d^* }} = \frac{{4\Phi }}{{3\pi 
n_d^{*1/3} }} - \frac{{\Phi (1 - \gamma )}}{{4\pi (1 + 
\Phi )}}, \label{e19}
\end{equation}
we obtain
\begin{equation}
\frac{{d\ln n_e }}{{dy}} = \left[ {\frac{2}{3} - 
\frac{{n_d^{*1/3} (1 - \gamma )}}{{8(1 + \Phi )}}} 
\right]\frac{{d\ln n_d }}{{dy}}. \label{e20}
\end{equation}

It follows from (\ref{e16})--(\ref{e18}) that at $a_p = a$, 
$F_{\mathrm{drv}} = 0$, i.e., a homogeneous cloud is 
stationary, as it must. For $a_p > a$, $\kappa < 1$
 and $F_{\mathrm{drv}} > 0$
 if $d\ln n_e /dy < 0$. Thus, a large subsonic probe moves 
from the void center toward the outer boundary of a cloud. 
Such conditions are typical for the experiment,\cite{49} 
where an outward motion of a large probe particle was 
registered. On the contrary, for $a_p < a$, $\kappa > 1$
 and $F_{\mathrm{drv}} < 0$, i.e., a small subsonic probe 
must move toward the void center. This effect was 
observed experimentally in Refs.~\onlinecite{68,63}.

Thus, the driving force acting on a probe particle originates 
from the dependence of the ion mean free path on the 
particle number density (\ref{e4}). Disregarding this 
dependence vanishes $F_{\mathrm{drv}} $. In fact, 
$\lambda \to \lambda _a$ if $n_d \to 0$, which means that 
$n_e \to n_i $. According to (\ref{e5}) in this case, 
$\gamma \to 0$
 and $F_{\mathrm{drv}} \to 0$
 as it follows from (\ref{e16}) and (\ref{e17}). The result 
$F_{\mathrm{drv}} \equiv 0$
 can be directly obtained using IEOS Eq.~(12) in 
Ref.~\onlinecite{64}, which is a limit case of 
Eqs.~(\ref{e7}) and (\ref{e8}) for $\gamma \ll 1$.

\section{\label{s5}VELOCITY OF THE PROBE 
MOTION}

\begin{figure}
\includegraphics[width=10cm]{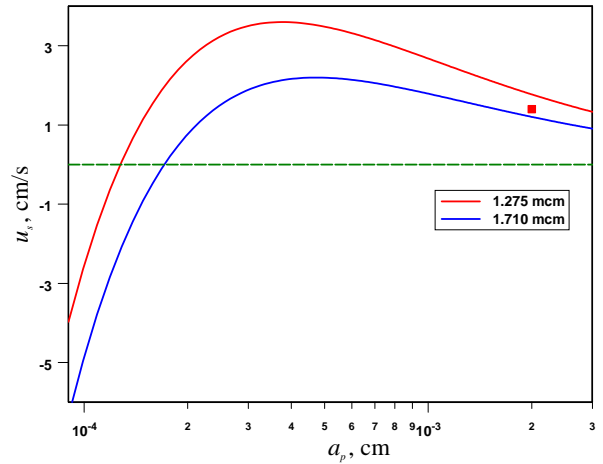}
\caption{\label{f4}Probe velocity as a function of its radius 
under conditions of the experiment\cite{49,53} for $a = 
1.275 \times 10^{ - 4}$ (red line) and $1.71 \times 10^{ - 
4} \;{\mbox{cm}}$
 (blue line). Red dot indicates the probe velocity from the 
experiment.\cite{49,53}}
\end{figure}

The Newtonian equation that governs the probe motion is
\begin{equation}
\dot u + \nu _p u = \frac{{F_{\mathrm{drv}} }}{{M_p }}, 
\label{e21}
\end{equation}
where $u$
 is the probe velocity, and
\begin{equation}
\nu _p = \frac{{8\sqrt {2\pi } }}{3}\frac{{\delta m_n n_n 
v_T a_p^2 }}{{M_p }}\left( {1 + \frac{1}{2}\sqrt 
{\frac{a}{{a_p }}} } \right) \label{e22}
\end{equation}
is the friction coefficient for a probe, $\delta \simeq 1.44$
 is the accommodation coefficient;\cite{33} $m_n$ is the 
mass of a gas molecule; $n_n$ and $v_T = (T_n /m_n 
)^{1/2}$ are the number density and thermal velocity of the 
gas molecules, respectively, $T_n = 300\;{\mbox{K}}$
 is the temperature of a gas; $M_p = (4\pi /3)\rho _p a_p^3$ 
and $\rho _p$ are the probe particle mass and its material 
density, respectively. Equation (\ref{e22}) allows for the 
enhancement of neutral drag force due to the dissipation in 
the fluid of dust particles surrounding the probe. It was 
shown in Ref.~\onlinecite{52} that this effect leads to an 
additional factor $1 + (R_p /r_d )^3 (a/a_p )^2$ in the 
expression for $\nu _p $. With due regard for 
Eq.~(\ref{e003}), we can rewrite this factor in the form that 
appears in (\ref{e22}). Note that if $a_p < a$, 
Eq.~(\ref{e22}) is invalid and it can only be regarded as an 
extrapolation. Since an appropriate estimation of the 
friction force for this case is absent, we will use (\ref{e22}) 
to plot Fig.~\ref{f4}. However, it is noteworthy that the 
effect of scattering of a smaller probe on larger particles 
forming the cloud can be more pronounced than the neutral 
drag---the probe drifts in strong fields of neighboring dust 
particles. Qualitatively, this is taken into account in 
(\ref{e22}).

The sustained probe velocity $u_s = F_{\mathrm{drv}} 
/\nu _p M_p$ calculated using Eqs.~(\ref{e16}), 
(\ref{e17}), (\ref{e20}), and (\ref{e22}) is shown in 
Fig.~\ref{f4} for two dust particle radii. For this estimation, 
we used the dust particle number density profiles 
determined recently for the foot region of the dust 
clouds.\cite{69} For these profiles, $\ln n_d /dy \approx - 
1.65\;{\mbox{cm}}^{ - 1}$ for $a = 1.275 \times 10^{ - 4} 
\;{\mbox{cm}}$
 and $\ln n_d /dy \approx - 1.87\;{\mbox{cm}}^{ - 1}$ for 
$a = 1.71 \times 10^{ - 4} \;{\mbox{cm}}$
 (processing these data assumed that $\beta = 1$), see 
Figs.~5 and 6 of Ref.~\onlinecite{69}, respectively. The 
electron temperature amounts to $3.5\;{\mbox{eV}}$. As 
$a_p$ is increased, the probe velocity changes its sign at 
$a_p = a$. It can be readily deduced from (\ref{e17}) and 
(\ref{e22}) that $u_s \propto 1/a_p$ at $a_p \to \infty $. 
Hence, $u_s (a_p )$
 has a maximum whose location depends on $a$
 (Fig.~\ref{f4}). In the experiment,\cite{49,53} the probe 
radius was assumed to be $a_p = 7.5 \times 10^{ - 4} 
\;{\mbox{cm}}$
 and its velocity reached $1.4\;{\mbox{cm/s}}$, which is of 
the same order of magnitude as in Fig.~\ref{f4} ($u_s 
\simeq 3.0\;{\mbox{cm/s}}$). Note that the radius of a 
probe particle could be underestimated in 
Ref.~\onlinecite{49}. Thus, at $a_p = 2.0 \times 10^{ - 3} 
\;{\mbox{cm}}$, $u_s \simeq 1.77\;{\mbox{cm/s}}$. Note 
that for this probe radius, the velocity relaxation time $\nu 
_p^{ - 1} \simeq 0.3\;{\mbox{s}}$
 is close to the total time of probe motion in a dust cloud 
($0.4\;{\mbox{s}}$), during which the sustained velocity 
cannot be attained, i.e., the effect of inertia can be essential. 
The difference between the theoretical estimate and 
experiment may also arise from a significant error of 
Eq.~(\ref{e2}) at high ion drift velocity comparable with 
their thermal velocity, which is characteristic of 
experiment.\cite{53} Another source of error could be the 
neglect of the mechanisms of particle charging other than 
the OML approximation such as the ion-neutral 
collisions.\cite{60,43} It was demonstrated that neglect of 
the collision effect overestimates somewhat the particle 
charge but this effect is small at the gas pressure less than 
$30\;{\mbox{Pa}}$
 and $Zn_d /n_e > 1$. Such conditions are typical for the 
experiments treated in this work. Note that since both SP 
and MSP neglect the dependence of $\sigma 
_{\mathrm{eff}}$ on the ion energy, inclusion of the 
ion-particle collisions in the calculation of the ion mean 
free path does not change the particle charge. Hence, the 
allowance for the effect of ion-neutral collisions would lead 
to an insignificant change of the results.

The discussion above implied that a probe is subsonic and a 
spherical cavity is formed around it as it moves through the 
dust cloud. In contrast, for a supersonic probe, a cavity is 
formed behind rather than around the probe, as it is 
illustrated with Fig.~\ref{f5} (see, e.g., 
Ref.~\onlinecite{19}). Parameters of such cavity are 
discussed in the Appendix~\ref{sA}. For a very crude 
estimation, one can assume that in the case of a large probe 
($a_p \gg a$), the ion-probe collision cross section is 
limited by that for a dust particle. Then 
$F_{\mathrm{drv}} \simeq (Z_p - Z)eE \simeq Z_p T_e 
(d\ln n_e /dy)$
 almost coincides with the electric force. Since typically 
$d\ln n_e /dy < 0$, a supersonic probe is driven toward the 
discharge center, i.e., in the opposite direction as compared 
to a subsonic probe. This effect was observed in 
experiment,\cite{19} where the probe velocity varied from 
$8$
 to $3.7\;{\mbox{cm/s}}$.
\begin{figure}
\includegraphics[width=8.7cm]{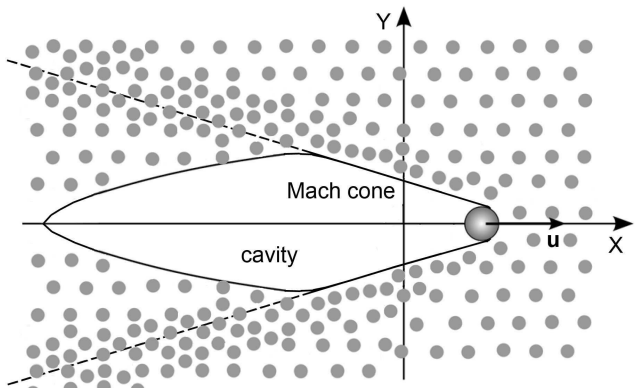}
\caption{\label{f5}Sketch showing a subsonic projectile 
(large bullet point) moving with the velocity $u$
 along the $X$-axis in a particle fluid (small bullet points 
around), with a neutral gas as a still background. The solid 
line indicates the boundary of a cavity and dashed line, the 
Mach cone that is the surface of particle number density 
perturbation.}
\end{figure}

\section{\label{s6} CONCLUSION}

In this work, we have calculated the sum of the ion drag 
force and the electric force (driving force) acting on a probe 
particle that finds itself in a dust cloud formed by the 
particles of the radius different from that of a probe in 
complex plasma of the low-pressure radio frequency 
discharge under microgravity conditions. We have modified 
our approach developed recently to include the effect of the 
ion collisions against the particles forming the cloud. This 
effect results in shortening the ion mean free path and, 
consequently, in the decrease of the moment transfer from 
the ions to the dust particles, i.e., to the decrease of the ion 
drag force. However, we have shown that the modified 
IEOS can be written in the form proposed in the early 
study.\cite{64} At the same time, the modified IEOS 
extends the range of plasma parameters considerably. Thus, 
it shifts the IEOS critical point toward much higher number 
densities of the ions, electrons, and particles. Fortunately, 
the velocity of dust acoustic waves calculated with the 
modified IEOS has a minor difference from that calculated 
with the initial version of IEOS and therefore, it still 
matches the experimental data.

In contrast to our early model (SP), in which the driving 
force vanishes, the modified model (MSP) yields a nonzero 
force, whose direction is uniquely defined by the ratio 
between the radius of the probe and of the dust particle. 
Namely, if the probe is smaller than a dust particle then the 
probe moves toward the discharge center against the ion 
flux, and the large probe moves in the same direction as the 
ions. Hence, obtained results provide an interpretation for 
the regularities observed in experiments where subsonic 
probes were studied. Estimation of the probe typical 
velocity leads to a reasonable correspondence with 
available experimental data. The developed model and the 
driving force calculated on its basis can serve as the ansatz 
for future investigation of such nonequilibrium phase 
transition in three-dimensional complex plasmas as the lane 
formation.

\begin{acknowledgments}
This research is supported by the Russian Science 
Foundation Grant No.~14-50-00124.
\end{acknowledgments}
\appendix
\section{\label{sA}PARAMETERS OF A CAVITY 
BEHIND THE SUPERSONIC PROBE}

We will estimate the maximum cavity width and the length 
of a cavity behind a supersonic probe moving with the 
velocity $u$. If $u \gg c_s$ then the Mach cone with 
contact discontinuity at its surface is realized 
(Fig.~\ref{f5}). If the probe is large, $M_d \ll M_p $, the 
velocity modulus of the dust particle scattered on the probe 
preserves. In this case, the transversal velocity component 
of the scattered dust particle is $u_ \bot  = u\sin \alpha = 
u/M = c_s $, where $\alpha$ is the Mach angle and $M = 
u/c_s$ is the Mach number. The expansion time is $1/\nu $, 
where $\nu = (8\sqrt {2\pi } /3)(\delta m_n n_n v_T a^2 
/M_d )$
 is the friction coefficient for a dust particle. Hence, the 
transversal dust particle displacement that coincides with 
the maximum cavity width is $\rho = c_s /\nu $. The Mach 
cone expansion length is then $l_e = \rho /\tan \alpha 
\simeq \rho M = u/\nu $.

Closing of the cavity is defined by the particle number 
density relaxation. Equation~(30) in Ref.~\onlinecite{64} is 
appropriate for the evolution of a density perturbation in 
complex plasmas. For a low-frequency perturbation, the 
overdamped regime takes place, and this equation assumes 
the form
\begin{equation}
\frac{{\partial \psi }}{{\partial t}} = \frac{{c_s^2 }}{\nu 
}\Delta \psi , \label{e091}
\end{equation}
where ${\bf{v}} = \bm{\nabla}\psi$ is the local dust 
particle velocity. It follows from (\ref{e091}) that the 
closing time and the corresponding closing length are $\tau 
_c = \nu \rho ^2 /c_s^2$ and $l_c = \nu M\rho ^2 /c_s $, 
respectively. The total cavity length $l = l_e + l_c$ is then
\begin{equation}
l = 2\rho M = \frac{{2u}}{\nu }. \label{e092}
\end{equation}
Note that within the accuracy of these estimates, $l_e$ 
proves to be equal to $l_c $. Since the damping frequency 
of dust acoustic waves is $\nu /2$,\cite{64} the length of 
Mach cone axis, $2u/\nu $, must coincide with the total 
cavity length (\ref{e092}).

Due to a strong decrease of the Coulomb momentum 
transfer cross section, an ultrafast probe can move through 
a dust cloud almost without perturbation of the cloud, i.e., 
without a cavity behind it. The threshold velocity of a probe 
can be estimated if we assume that the transversal particle 
displacement is equal to $r_d /2$. Obviously, the typical 
impact parameter is the same, $\rho _0 \simeq r_d /2$. For 
the transversal direction, the dust particle equation of 
motion has the form
\begin{equation}
\ddot z + \nu \dot z = \frac{{aa_p \Phi ^2 T_e^2 }}{{M_d 
e^2 }}\frac{1}{{ut + \rho _0^2 }}. \label{e093}
\end{equation}
At the first stage, the particle gains momentum from a 
probe, $\left| {\ddot z} \right| \gg \nu \left| {\dot z} \right|$, 
and its displacement is small. At the second stage, the 
particle is decelerated due to the neutral drag, $\ddot z + \nu 
\dot z \simeq 0$. Calculation of the transversal 
displacement makes it possible to obtain the estimate for 
threshold velocity
\begin{equation}
u_{\mathrm{th}} = \frac{{8aa_p \Phi ^2 T_e^2 }}{{\nu 
M_d e^2 r_d^2 }}. \label{e094}
\end{equation}
For $u > u_{\mathrm{th}} $, the cavity does not open.

We will estimate the calculated parameters of a cavity 
behind a supersonic probe for the experiment,\cite{11} 
where $a = 4.775 \times 10^{ - 4} \;{\mbox{cm}}$, $M_d 
= 6.89 \times 10^{ - 10} \;{\mbox{g}}$, $r_d = 2.29 \times 
10^{ - 2} \;{\mbox{cm}}$, $T_e = 4.5\;{\mbox{eV}}$, 
$\Phi = 1.8$, $c_s = 2.0\;{\mbox{cm/s}}$, $\nu = 
33.5\;{\mbox{s}}^{ - 1} $, $a_p = 10^{ - 3} 
\;{\mbox{cm}}$, and $u = 13.2\;{\mbox{cm/s}}$
 (${\mbox{M}} = 6.6$). Under these conditions, we obtain 
the maximum cavity width $\rho \simeq 
0.06\;{\mbox{cm}}$, its expansion velocity $u_ \bot  = 
2.0\;{\mbox{cm/s}}$, and its total length $l \simeq 
0.79\;{\mbox{cm}}$
 (Eq.~(\ref{e092})). The experimental parameters are 
$0.05\;{\mbox{cm}}$, $2.1\;{\mbox{cm/s}}$, and 
$1.1\;{\mbox{cm}}$, respectively. One can testify a 
satisfactory correspondence between these estimates and 
the experiment. From (\ref{e094}), we obtain the threshold 
velocity $u_{\mathrm{th}} = 232\;{\mbox{cm/s}}$, which 
also correlates with the experimental assessment 
($u_{\mathrm{th}} = 158\;{\mbox{cm/s}}$). Consider the 
experiment,\cite{19} for which $\nu = 49\;{\mbox{s}}^{ - 
1} $, $c_s = 2.8\;{\mbox{cm/s}}$, and $u = 
6\;{\mbox{cm/s}}$. Our estimations yield $\rho \simeq 
0.057\;{\mbox{cm}}$
 and $l \simeq 0.25\;{\mbox{cm}}$
 vs.\ $0.045\;{\mbox{cm}}$
 and $0.31\;{\mbox{cm}}$
 from the experiment.

\providecommand{\noopsort}[1]{}\providecommand{\singleletter}[1]{#1}%


\begin{thebibliography}{44}%
\makeatletter
\providecommand \@ifxundefined [1]{%
 \@ifx{#1\undefined}
}%
\providecommand \@ifnum [1]{%
 \ifnum #1\expandafter \@firstoftwo
 \else \expandafter \@secondoftwo
 \fi
}%
\providecommand \@ifx [1]{%
 \ifx #1\expandafter \@firstoftwo
 \else \expandafter \@secondoftwo
 \fi
}%
\providecommand \natexlab [1]{#1}%
\providecommand \enquote  [1]{``#1''}%
\providecommand \bibnamefont  [1]{#1}%
\providecommand \bibfnamefont [1]{#1}%
\providecommand \citenamefont [1]{#1}%
\providecommand \href@noop [0]{\@secondoftwo}%
\providecommand \href [0]{\begingroup \@sanitize@url \@href}%
\providecommand \@href[1]{\@@startlink{#1}\@@href}%
\providecommand \@@href[1]{\endgroup#1\@@endlink}%
\providecommand \@sanitize@url [0]{\catcode `\\12\catcode `\$12\catcode
  `\&12\catcode `\#12\catcode `\^12\catcode `\_12\catcode `\%12\relax}%
\providecommand \@@startlink[1]{}%
\providecommand \@@endlink[0]{}%
\providecommand \url  [0]{\begingroup\@sanitize@url \@url }%
\providecommand \@url [1]{\endgroup\@href {#1}{\urlprefix }}%
\providecommand \urlprefix  [0]{URL }%
\providecommand \Eprint [0]{\href }%
\providecommand \doibase [0]{http://dx.doi.org/}%
\providecommand \selectlanguage [0]{\@gobble}%
\providecommand \bibinfo  [0]{\@secondoftwo}%
\providecommand \bibfield  [0]{\@secondoftwo}%
\providecommand \translation [1]{[#1]}%
\providecommand \BibitemOpen [0]{}%
\providecommand \bibitemStop [0]{}%
\providecommand \bibitemNoStop [0]{.\EOS\space}%
\providecommand \EOS [0]{\spacefactor3000\relax}%
\providecommand \BibitemShut  [1]{\csname bibitem#1\endcsname}%
\let\auto@bib@innerbib\@empty
\bibitem [{\citenamefont {Morfill}\ \emph {et~al.}(2006)\citenamefont
  {Morfill}, \citenamefont {Konopka}, \citenamefont {Kretschmer}, \citenamefont
  {Rubin-Zuzic}, \citenamefont {Thomas}, \citenamefont {Zhdanov},\ and\
  \citenamefont {Tsytovich}}]{10}%
  \BibitemOpen
  \bibfield  {author} {\bibinfo {author} {\bibfnamefont {G.~E.}\ \bibnamefont
  {Morfill}}, \bibinfo {author} {\bibfnamefont {U.}~\bibnamefont {Konopka}},
  \bibinfo {author} {\bibfnamefont {M.}~\bibnamefont {Kretschmer}}, \bibinfo
  {author} {\bibfnamefont {M.}~\bibnamefont {Rubin-Zuzic}}, \bibinfo {author}
  {\bibfnamefont {H.~M.}\ \bibnamefont {Thomas}}, \bibinfo {author}
  {\bibfnamefont {S.~K.}\ \bibnamefont {Zhdanov}}, \ and\ \bibinfo {author}
  {\bibfnamefont {V.}~\bibnamefont {Tsytovich}},\ }\href {\doibase
  10.1088/1367-2630/8/1/007} {\bibfield  {journal} {\bibinfo  {journal} {New
  J.\ Phys.}\ }\textbf {\bibinfo {volume} {8}},\ \bibinfo {pages} {7} (\bibinfo
  {year} {2006})}\BibitemShut {NoStop}%
\bibitem [{\citenamefont {Caliebe}\ \emph {et~al.}(2011)\citenamefont
  {Caliebe}, \citenamefont {Arp},\ and\ \citenamefont {Piel}}]{11}%
  \BibitemOpen
  \bibfield  {author} {\bibinfo {author} {\bibfnamefont {D.}~\bibnamefont
  {Caliebe}}, \bibinfo {author} {\bibfnamefont {O.}~\bibnamefont {Arp}}, \ and\
  \bibinfo {author} {\bibfnamefont {A.}~\bibnamefont {Piel}},\ }\href {\doibase
  10.1063/1.3606468} {\bibfield  {journal} {\bibinfo  {journal} {Phys.\
  Plasmas}\ }\textbf {\bibinfo {volume} {18}},\ \bibinfo {pages} {073702}
  (\bibinfo {year} {2011})}\BibitemShut {NoStop}%
\bibitem [{\citenamefont {Piel}\ \emph {et~al.}(2008)\citenamefont {Piel},
  \citenamefont {Arp}, \citenamefont {Klindworth},\ and\ \citenamefont
  {Melzer}}]{12}%
  \BibitemOpen
  \bibfield  {author} {\bibinfo {author} {\bibfnamefont {A.}~\bibnamefont
  {Piel}}, \bibinfo {author} {\bibfnamefont {O.}~\bibnamefont {Arp}}, \bibinfo
  {author} {\bibfnamefont {M.}~\bibnamefont {Klindworth}}, \ and\ \bibinfo
  {author} {\bibfnamefont {A.}~\bibnamefont {Melzer}},\ }\href {\doibase
  10.1103/PhysRevE.77.026407} {\bibfield  {journal} {\bibinfo  {journal}
  {Phys.\ Rev.\ E}\ }\textbf {\bibinfo {volume} {77}},\ \bibinfo {pages}
  {026407} (\bibinfo {year} {2008})}\BibitemShut {NoStop}%
\bibitem [{\citenamefont {Menzel}\ \emph {et~al.}(2011)\citenamefont {Menzel},
  \citenamefont {Arp},\ and\ \citenamefont {Piel}}]{13}%
  \BibitemOpen
  \bibfield  {author} {\bibinfo {author} {\bibfnamefont {K.~O.}\ \bibnamefont
  {Menzel}}, \bibinfo {author} {\bibfnamefont {O.}~\bibnamefont {Arp}}, \ and\
  \bibinfo {author} {\bibfnamefont {A.}~\bibnamefont {Piel}},\ }\href {\doibase
  10.1103/PhysRevE.83.016402} {\bibfield  {journal} {\bibinfo  {journal}
  {Phys.\ Rev.\ E}\ }\textbf {\bibinfo {volume} {83}},\ \bibinfo {pages}
  {016402} (\bibinfo {year} {2011})}\BibitemShut {NoStop}%
\bibitem [{\citenamefont {Arp}\ \emph {et~al.}(2011)\citenamefont {Arp},
  \citenamefont {Caliebe},\ and\ \citenamefont {Piel}}]{14}%
  \BibitemOpen
  \bibfield  {author} {\bibinfo {author} {\bibfnamefont {O.}~\bibnamefont
  {Arp}}, \bibinfo {author} {\bibfnamefont {D.}~\bibnamefont {Caliebe}}, \ and\
  \bibinfo {author} {\bibfnamefont {A.}~\bibnamefont {Piel}},\ }\href {\doibase
  10.1103/PhysRevE.83.066404} {\bibfield  {journal} {\bibinfo  {journal}
  {Phys.\ Rev.\ E}\ }\textbf {\bibinfo {volume} {83}},\ \bibinfo {pages}
  {066404} (\bibinfo {year} {2011})}\BibitemShut {NoStop}%
\bibitem [{\citenamefont {Schwabe}\ \emph {et~al.}(2008)\citenamefont
  {Schwabe}, \citenamefont {Zhdanov}, \citenamefont {Thomas}, \citenamefont
  {Ivlev}, \citenamefont {Rubin-Zuzic}, \citenamefont {Morfill}, \citenamefont
  {Molotkov}, \citenamefont {Lipaev}, \citenamefont {Fortov},\ and\
  \citenamefont {Reiter}}]{15}%
  \BibitemOpen
  \bibfield  {author} {\bibinfo {author} {\bibfnamefont {M.}~\bibnamefont
  {Schwabe}}, \bibinfo {author} {\bibfnamefont {S.~K.}\ \bibnamefont
  {Zhdanov}}, \bibinfo {author} {\bibfnamefont {H.~M.}\ \bibnamefont {Thomas}},
  \bibinfo {author} {\bibfnamefont {A.~V.}\ \bibnamefont {Ivlev}}, \bibinfo
  {author} {\bibfnamefont {M.}~\bibnamefont {Rubin-Zuzic}}, \bibinfo {author}
  {\bibfnamefont {G.~E.}\ \bibnamefont {Morfill}}, \bibinfo {author}
  {\bibfnamefont {V.~I.}\ \bibnamefont {Molotkov}}, \bibinfo {author}
  {\bibfnamefont {A.~M.}\ \bibnamefont {Lipaev}}, \bibinfo {author}
  {\bibfnamefont {V.~E.}\ \bibnamefont {Fortov}}, \ and\ \bibinfo {author}
  {\bibfnamefont {T.}~\bibnamefont {Reiter}},\ }\href {\doibase
  10.1088/1367-2630/10/3/033037} {\bibfield  {journal} {\bibinfo  {journal}
  {New J.\ Phys.}\ }\textbf {\bibinfo {volume} {10}},\ \bibinfo {pages}
  {033037} (\bibinfo {year} {2008})}\BibitemShut {NoStop}%
\bibitem [{\citenamefont {Morfill}\ \emph {et~al.}(1999)\citenamefont
  {Morfill}, \citenamefont {Thomas}, \citenamefont {Konopka}, \citenamefont
  {Rothermel}, \citenamefont {Zuzic}, \citenamefont {Ivlev},\ and\
  \citenamefont {Goree}}]{16}%
  \BibitemOpen
  \bibfield  {author} {\bibinfo {author} {\bibfnamefont {G.~E.}\ \bibnamefont
  {Morfill}}, \bibinfo {author} {\bibfnamefont {H.~M.}\ \bibnamefont {Thomas}},
  \bibinfo {author} {\bibfnamefont {U.}~\bibnamefont {Konopka}}, \bibinfo
  {author} {\bibfnamefont {H.}~\bibnamefont {Rothermel}}, \bibinfo {author}
  {\bibfnamefont {M.}~\bibnamefont {Zuzic}}, \bibinfo {author} {\bibfnamefont
  {A.}~\bibnamefont {Ivlev}}, \ and\ \bibinfo {author} {\bibfnamefont
  {J.}~\bibnamefont {Goree}},\ }\href {\doibase 10.1103/PhysRevLett.83.1598}
  {\bibfield  {journal} {\bibinfo  {journal} {Phys.\ Rev.\ Lett.}\ }\textbf
  {\bibinfo {volume} {83}},\ \bibinfo {pages} {1598} (\bibinfo {year}
  {1999})}\BibitemShut {NoStop}%
\bibitem [{\citenamefont {Khrapak}\ \emph {et~al.}(2011)\citenamefont
  {Khrapak}, \citenamefont {Klumov}, \citenamefont {Huber}, \citenamefont
  {Molotkov}, \citenamefont {Lipaev}, \citenamefont {Naumkin}, \citenamefont
  {Thomas}, \citenamefont {Ivlev}, \citenamefont {Morfill}, \citenamefont
  {Petrov}, \citenamefont {Fortov}, \citenamefont {Malentschenko},\ and\
  \citenamefont {Volkov}}]{17}%
  \BibitemOpen
  \bibfield  {author} {\bibinfo {author} {\bibfnamefont {S.~A.}\ \bibnamefont
  {Khrapak}}, \bibinfo {author} {\bibfnamefont {B.~A.}\ \bibnamefont {Klumov}},
  \bibinfo {author} {\bibfnamefont {P.}~\bibnamefont {Huber}}, \bibinfo
  {author} {\bibfnamefont {V.~I.}\ \bibnamefont {Molotkov}}, \bibinfo {author}
  {\bibfnamefont {A.~M.}\ \bibnamefont {Lipaev}}, \bibinfo {author}
  {\bibfnamefont {V.~N.}\ \bibnamefont {Naumkin}}, \bibinfo {author}
  {\bibfnamefont {H.~M.}\ \bibnamefont {Thomas}}, \bibinfo {author}
  {\bibfnamefont {A.~V.}\ \bibnamefont {Ivlev}}, \bibinfo {author}
  {\bibfnamefont {G.~E.}\ \bibnamefont {Morfill}}, \bibinfo {author}
  {\bibfnamefont {O.~F.}\ \bibnamefont {Petrov}}, \bibinfo {author}
  {\bibfnamefont {V.~E.}\ \bibnamefont {Fortov}}, \bibinfo {author}
  {\bibfnamefont {{\relax Yu}.}~\bibnamefont {Malentschenko}}, \ and\ \bibinfo
  {author} {\bibfnamefont {S.}~\bibnamefont {Volkov}},\ }\href {\doibase
  10.1103/PhysRevLett.106.205001} {\bibfield  {journal} {\bibinfo  {journal}
  {Phys.\ Rev.\ Lett.}\ }\textbf {\bibinfo {volume} {106}},\ \bibinfo {pages}
  {205001} (\bibinfo {year} {2011})}\BibitemShut {NoStop}%
\bibitem [{\citenamefont {Thomas}\ \emph {et~al.}(2008)\citenamefont {Thomas},
  \citenamefont {Morfill}, \citenamefont {Fortov}, \citenamefont {Ivlev},
  \citenamefont {Molotkov}, \citenamefont {Lipaev}, \citenamefont {Hagl},
  \citenamefont {Rothermel}, \citenamefont {Khrapak}, \citenamefont
  {Suetterlin}, \citenamefont {Rubin-Zuzic}, \citenamefont {Petrov},
  \citenamefont {Tokarev},\ and\ \citenamefont {Krikalev}}]{18}%
  \BibitemOpen
  \bibfield  {author} {\bibinfo {author} {\bibfnamefont {H.~M.}\ \bibnamefont
  {Thomas}}, \bibinfo {author} {\bibfnamefont {G.~E.}\ \bibnamefont {Morfill}},
  \bibinfo {author} {\bibfnamefont {V.~E.}\ \bibnamefont {Fortov}}, \bibinfo
  {author} {\bibfnamefont {A.~V.}\ \bibnamefont {Ivlev}}, \bibinfo {author}
  {\bibfnamefont {V.~I.}\ \bibnamefont {Molotkov}}, \bibinfo {author}
  {\bibfnamefont {A.~M.}\ \bibnamefont {Lipaev}}, \bibinfo {author}
  {\bibfnamefont {T.}~\bibnamefont {Hagl}}, \bibinfo {author} {\bibfnamefont
  {H.}~\bibnamefont {Rothermel}}, \bibinfo {author} {\bibfnamefont {S.~A.}\
  \bibnamefont {Khrapak}}, \bibinfo {author} {\bibfnamefont {R.~K.}\
  \bibnamefont {Suetterlin}}, \bibinfo {author} {\bibfnamefont
  {M.}~\bibnamefont {Rubin-Zuzic}}, \bibinfo {author} {\bibfnamefont {O.~F.}\
  \bibnamefont {Petrov}}, \bibinfo {author} {\bibfnamefont {V.~I.}\
  \bibnamefont {Tokarev}}, \ and\ \bibinfo {author} {\bibfnamefont {S.~K.}\
  \bibnamefont {Krikalev}},\ }\href {\doibase 10.1088/1367-2630/10/3/033036}
  {\bibfield  {journal} {\bibinfo  {journal} {New J.\ Phys.}\ }\textbf
  {\bibinfo {volume} {10}},\ \bibinfo {pages} {033036} (\bibinfo {year}
  {2008})}\BibitemShut {NoStop}%
\bibitem [{\citenamefont {Jiang}\ \emph {et~al.}(2009)\citenamefont {Jiang},
  \citenamefont {Nosenko}, \citenamefont {Li}, \citenamefont {Schwabe},
  \citenamefont {Konopka}, \citenamefont {Ivlev}, \citenamefont {Fortov},
  \citenamefont {Molotkov}, \citenamefont {Lipaev}, \citenamefont {Petrov},
  \citenamefont {Turin}, \citenamefont {Thomas},\ and\ \citenamefont
  {Morfill}}]{019}%
  \BibitemOpen
  \bibfield  {author} {\bibinfo {author} {\bibfnamefont {K.}~\bibnamefont
  {Jiang}}, \bibinfo {author} {\bibfnamefont {V.}~\bibnamefont {Nosenko}},
  \bibinfo {author} {\bibfnamefont {Y.~F.}\ \bibnamefont {Li}}, \bibinfo
  {author} {\bibfnamefont {M.}~\bibnamefont {Schwabe}}, \bibinfo {author}
  {\bibfnamefont {U.}~\bibnamefont {Konopka}}, \bibinfo {author} {\bibfnamefont
  {A.~V.}\ \bibnamefont {Ivlev}}, \bibinfo {author} {\bibfnamefont {V.~E.}\
  \bibnamefont {Fortov}}, \bibinfo {author} {\bibfnamefont {V.~I.}\
  \bibnamefont {Molotkov}}, \bibinfo {author} {\bibfnamefont {A.~M.}\
  \bibnamefont {Lipaev}}, \bibinfo {author} {\bibfnamefont {O.~F.}\
  \bibnamefont {Petrov}}, \bibinfo {author} {\bibfnamefont {M.~V.}\
  \bibnamefont {Turin}}, \bibinfo {author} {\bibfnamefont {H.~M.}\ \bibnamefont
  {Thomas}}, \ and\ \bibinfo {author} {\bibfnamefont {G.~E.}\ \bibnamefont
  {Morfill}},\ }\href {\doibase 10.1209/0295-5075/85/45002} {\bibfield
  {journal} {\bibinfo  {journal} {Europhys.\ Lett.}\ }\textbf {\bibinfo
  {volume} {85}},\ \bibinfo {pages} {45002} (\bibinfo {year}
  {2009})}\BibitemShut {NoStop}%
\bibitem [{\citenamefont {Schwabe}\ \emph {et~al.}(2011)\citenamefont
  {Schwabe}, \citenamefont {Jiang}, \citenamefont {Zhdanov}, \citenamefont
  {Hagl}, \citenamefont {Huber}, \citenamefont {Ivlev}, \citenamefont {Lipaev},
  \citenamefont {Molotkov}, \citenamefont {Naumkin}, \citenamefont
  {S{\"{u}}tterlin}, \citenamefont {Thomas}, \citenamefont {Fortov},
  \citenamefont {Morfill}, \citenamefont {Skvortsov},\ and\ \citenamefont
  {Volkov}}]{19}%
  \BibitemOpen
  \bibfield  {author} {\bibinfo {author} {\bibfnamefont {M.}~\bibnamefont
  {Schwabe}}, \bibinfo {author} {\bibfnamefont {K.}~\bibnamefont {Jiang}},
  \bibinfo {author} {\bibfnamefont {S.}~\bibnamefont {Zhdanov}}, \bibinfo
  {author} {\bibfnamefont {T.}~\bibnamefont {Hagl}}, \bibinfo {author}
  {\bibfnamefont {P.}~\bibnamefont {Huber}}, \bibinfo {author} {\bibfnamefont
  {A.~V.}\ \bibnamefont {Ivlev}}, \bibinfo {author} {\bibfnamefont {A.~M.}\
  \bibnamefont {Lipaev}}, \bibinfo {author} {\bibfnamefont {V.~I.}\
  \bibnamefont {Molotkov}}, \bibinfo {author} {\bibfnamefont {V.~N.}\
  \bibnamefont {Naumkin}}, \bibinfo {author} {\bibfnamefont {K.~R.}\
  \bibnamefont {S{\"{u}}tterlin}}, \bibinfo {author} {\bibfnamefont {H.~M.}\
  \bibnamefont {Thomas}}, \bibinfo {author} {\bibfnamefont {V.~E.}\
  \bibnamefont {Fortov}}, \bibinfo {author} {\bibfnamefont {G.~E.}\
  \bibnamefont {Morfill}}, \bibinfo {author} {\bibfnamefont {A.}~\bibnamefont
  {Skvortsov}}, \ and\ \bibinfo {author} {\bibfnamefont {S.}~\bibnamefont
  {Volkov}},\ }\href {\doibase 10.1209/0295-5075/96/55001} {\bibfield
  {journal} {\bibinfo  {journal} {Europhys.\ Lett.}\ }\textbf {\bibinfo
  {volume} {96}},\ \bibinfo {pages} {55001} (\bibinfo {year}
  {2011})}\BibitemShut {NoStop}%
\bibitem [{\citenamefont {Fortov}\ and\ \citenamefont {Morfill}(2010)}]{1}%
  \BibitemOpen
  \bibinfo {editor} {\bibfnamefont {V.~E.}\ \bibnamefont {Fortov}}\ and\
  \bibinfo {editor} {\bibfnamefont {G.~E.}\ \bibnamefont {Morfill}},\ eds.,\
  \href@noop {} {\emph {\bibinfo {title} {Complex and Dusty Plasmas: From
  Laboratory to Space}}},\ Series in Plasma Physics\ (\bibinfo  {publisher}
  {CRC Press},\ \bibinfo {address} {Boca Raton, FL},\ \bibinfo {year}
  {2010})\BibitemShut {NoStop}%
\bibitem [{\citenamefont {Chu}\ and\ \citenamefont {{Lin I}}(1994)}]{2}%
  \BibitemOpen
  \bibfield  {author} {\bibinfo {author} {\bibfnamefont {J.~H.}\ \bibnamefont
  {Chu}}\ and\ \bibinfo {author} {\bibnamefont {{Lin I}}},\ }\href {\doibase
  10.1103/PhysRevLett.72.4009} {\bibfield  {journal} {\bibinfo  {journal}
  {Phys.\ Rev.\ Lett.}\ }\textbf {\bibinfo {volume} {72}},\ \bibinfo {pages}
  {4009} (\bibinfo {year} {1994})}\BibitemShut {NoStop}%
\bibitem [{\citenamefont {Thomas}\ \emph {et~al.}(1994)\citenamefont {Thomas},
  \citenamefont {Morfill}, \citenamefont {Demmel}, \citenamefont {Goree},
  \citenamefont {Feuerbacher},\ and\ \citenamefont {M{\"{o}}hlmann}}]{3}%
  \BibitemOpen
  \bibfield  {author} {\bibinfo {author} {\bibfnamefont {H.}~\bibnamefont
  {Thomas}}, \bibinfo {author} {\bibfnamefont {G.~E.}\ \bibnamefont {Morfill}},
  \bibinfo {author} {\bibfnamefont {V.}~\bibnamefont {Demmel}}, \bibinfo
  {author} {\bibfnamefont {J.}~\bibnamefont {Goree}}, \bibinfo {author}
  {\bibfnamefont {B.}~\bibnamefont {Feuerbacher}}, \ and\ \bibinfo {author}
  {\bibfnamefont {D.}~\bibnamefont {M{\"{o}}hlmann}},\ }\href {\doibase
  10.1103/PhysRevLett.73.652} {\bibfield  {journal} {\bibinfo  {journal}
  {Phys.\ Rev.\ Lett.}\ }\textbf {\bibinfo {volume} {73}},\ \bibinfo {pages}
  {652} (\bibinfo {year} {1994})}\BibitemShut {NoStop}%
\bibitem [{\citenamefont {Hayashi}\ and\ \citenamefont {Tashibana}(1994)}]{4}%
  \BibitemOpen
  \bibfield  {author} {\bibinfo {author} {\bibfnamefont {Y.}~\bibnamefont
  {Hayashi}}\ and\ \bibinfo {author} {\bibfnamefont {S.}~\bibnamefont
  {Tashibana}},\ }\href {\doibase 10.1143/JJAP.33.L804} {\bibfield  {journal}
  {\bibinfo  {journal} {Jpn.\ J.\ Appl.\ Phys.}\ }\textbf {\bibinfo {volume}
  {33}},\ \bibinfo {pages} {L804} (\bibinfo {year} {1994})}\BibitemShut
  {NoStop}%
\bibitem [{\citenamefont {Vladimirov}\ \emph {et~al.}(2005)\citenamefont
  {Vladimirov}, \citenamefont {Ostrikov},\ and\ \citenamefont {Samarian}}]{5}%
  \BibitemOpen
  \bibfield  {author} {\bibinfo {author} {\bibfnamefont {S.~V.}\ \bibnamefont
  {Vladimirov}}, \bibinfo {author} {\bibfnamefont {K.}~\bibnamefont
  {Ostrikov}}, \ and\ \bibinfo {author} {\bibfnamefont {A.~A.}\ \bibnamefont
  {Samarian}},\ }\href@noop {} {\emph {\bibinfo {title} {Physics and
  Applications of Complex Plasmas}}}\ (\bibinfo  {publisher} {Imperial
  College},\ \bibinfo {address} {London},\ \bibinfo {year} {2005})\BibitemShut
  {NoStop}%
\bibitem [{\citenamefont {Fortov}\ \emph {et~al.}(2005)\citenamefont {Fortov},
  \citenamefont {Ivlev}, \citenamefont {Khrapak}, \citenamefont {Khrapak},\
  and\ \citenamefont {Morfill}}]{6}%
  \BibitemOpen
  \bibfield  {author} {\bibinfo {author} {\bibfnamefont {V.}~\bibnamefont
  {Fortov}}, \bibinfo {author} {\bibfnamefont {A.}~\bibnamefont {Ivlev}},
  \bibinfo {author} {\bibfnamefont {S.}~\bibnamefont {Khrapak}}, \bibinfo
  {author} {\bibfnamefont {A.}~\bibnamefont {Khrapak}}, \ and\ \bibinfo
  {author} {\bibfnamefont {G.}~\bibnamefont {Morfill}},\ }\href {\doibase
  10.1016/j.physrep.2005.08.007} {\bibfield  {journal} {\bibinfo  {journal}
  {Phys.\ Rep.}\ }\textbf {\bibinfo {volume} {421}},\ \bibinfo {pages} {1}
  (\bibinfo {year} {2005})}\BibitemShut {NoStop}%
\bibitem [{\citenamefont {Shukla}\ and\ \citenamefont {Eliasson}(2009)}]{8}%
  \BibitemOpen
  \bibfield  {author} {\bibinfo {author} {\bibfnamefont {P.~K.}\ \bibnamefont
  {Shukla}}\ and\ \bibinfo {author} {\bibfnamefont {B.}~\bibnamefont
  {Eliasson}},\ }\href {\doibase 10.1103/RevModPhys.81.25} {\bibfield
  {journal} {\bibinfo  {journal} {Rev.\ Mod.\ Phys.}\ }\textbf {\bibinfo
  {volume} {81}},\ \bibinfo {pages} {25} (\bibinfo {year} {2009})}\BibitemShut
  {NoStop}%
\bibitem [{\citenamefont {Bonitz}\ \emph {et~al.}(2010)\citenamefont {Bonitz},
  \citenamefont {Henning},\ and\ \citenamefont {Block}}]{9}%
  \BibitemOpen
  \bibfield  {author} {\bibinfo {author} {\bibfnamefont {M.}~\bibnamefont
  {Bonitz}}, \bibinfo {author} {\bibfnamefont {C.}~\bibnamefont {Henning}}, \
  and\ \bibinfo {author} {\bibfnamefont {D.}~\bibnamefont {Block}},\ }\href
  {\doibase 10.1088/0034-4885/73/6/066501} {\bibfield  {journal} {\bibinfo
  {journal} {Rep.\ Prog.\ Phys.}\ }\textbf {\bibinfo {volume} {73}},\ \bibinfo
  {pages} {066501} (\bibinfo {year} {2010})}\BibitemShut {NoStop}%
\bibitem [{\citenamefont {Schmittmann}\ and\ \citenamefont {Zia}(1998)}]{72}%
  \BibitemOpen
  \bibfield  {author} {\bibinfo {author} {\bibfnamefont {B.}~\bibnamefont
  {Schmittmann}}\ and\ \bibinfo {author} {\bibfnamefont {R.}~\bibnamefont
  {Zia}},\ }\href {\doibase 10.1016/S0370-1573(98)00005-2} {\bibfield
  {journal} {\bibinfo  {journal} {Phys.\ Rep.}\ }\textbf {\bibinfo {volume}
  {301}},\ \bibinfo {pages} {45} (\bibinfo {year} {1998})}\BibitemShut
  {NoStop}%
\bibitem [{\citenamefont {Ciamarra}\ \emph {et~al.}(2006)\citenamefont
  {Ciamarra}, \citenamefont {Coniglio},\ and\ \citenamefont {Nicodemi}}]{74}%
  \BibitemOpen
  \bibfield  {author} {\bibinfo {author} {\bibfnamefont {M.}~\bibnamefont
  {Ciamarra}}, \bibinfo {author} {\bibfnamefont {A.}~\bibnamefont {Coniglio}},
  \ and\ \bibinfo {author} {\bibfnamefont {M.}~\bibnamefont {Nicodemi}},\
  }\href {\doibase 10.1103/PhysRevLett.97.038001} {\bibfield  {journal}
  {\bibinfo  {journal} {Phys.\ Rev.\ Lett.}\ }\textbf {\bibinfo {volume}
  {97}},\ \bibinfo {pages} {038001} (\bibinfo {year} {2006})}\BibitemShut
  {NoStop}%
\bibitem [{\citenamefont {Netz}(2003)}]{83}%
  \BibitemOpen
  \bibfield  {author} {\bibinfo {author} {\bibfnamefont {R.~R.}\ \bibnamefont
  {Netz}},\ }\href {\doibase 10.1209/epl/i2003-00557-x} {\bibfield  {journal}
  {\bibinfo  {journal} {Europhys.\ Lett.}\ }\textbf {\bibinfo {volume} {63}},\
  \bibinfo {pages} {616} (\bibinfo {year} {2003})}\BibitemShut {NoStop}%
\bibitem [{\citenamefont {Helbing}\ \emph {et~al.}(2000)\citenamefont
  {Helbing}, \citenamefont {Farkas},\ and\ \citenamefont {Vicsek}}]{73}%
  \BibitemOpen
  \bibfield  {author} {\bibinfo {author} {\bibfnamefont {D.}~\bibnamefont
  {Helbing}}, \bibinfo {author} {\bibfnamefont {I.~J.}\ \bibnamefont {Farkas}},
  \ and\ \bibinfo {author} {\bibfnamefont {T.}~\bibnamefont {Vicsek}},\ }\href
  {\doibase 10.1103/PhysRevLett.84.1240} {\bibfield  {journal} {\bibinfo
  {journal} {Phys.\ Rev.\ Lett.}\ }\textbf {\bibinfo {volume} {84}},\ \bibinfo
  {pages} {1240} (\bibinfo {year} {2000})}\BibitemShut {NoStop}%
\bibitem [{\citenamefont {L{\"{o}}wen}\ and\ \citenamefont
  {Dzubiella}(2003)}]{75}%
  \BibitemOpen
  \bibfield  {author} {\bibinfo {author} {\bibfnamefont {H.}~\bibnamefont
  {L{\"{o}}wen}}\ and\ \bibinfo {author} {\bibfnamefont {J.}~\bibnamefont
  {Dzubiella}},\ }\href {\doibase 10.1039/b202892c} {\bibfield  {journal}
  {\bibinfo  {journal} {Faraday Discuss.}\ }\textbf {\bibinfo {volume} {123}},\
  \bibinfo {pages} {99} (\bibinfo {year} {2003})}\BibitemShut {NoStop}%
\bibitem [{\citenamefont {Dzubiella}\ \emph {et~al.}(2002)\citenamefont
  {Dzubiella}, \citenamefont {Hoffmann},\ and\ \citenamefont
  {L{\"{o}}wen}}]{76}%
  \BibitemOpen
  \bibfield  {author} {\bibinfo {author} {\bibfnamefont {J.}~\bibnamefont
  {Dzubiella}}, \bibinfo {author} {\bibfnamefont {G.~P.}\ \bibnamefont
  {Hoffmann}}, \ and\ \bibinfo {author} {\bibfnamefont {H.}~\bibnamefont
  {L{\"{o}}wen}},\ }\href {\doibase 10.1103/PhysRevE.65.021402} {\bibfield
  {journal} {\bibinfo  {journal} {Phys.\ Rev.\ E}\ }\textbf {\bibinfo {volume}
  {65}},\ \bibinfo {pages} {021402} (\bibinfo {year} {2002})}\BibitemShut
  {NoStop}%
\bibitem [{\citenamefont {Dzubiella}\ and\ \citenamefont
  {L{\"{o}}wen}(2002)}]{77}%
  \BibitemOpen
  \bibfield  {author} {\bibinfo {author} {\bibfnamefont {J.}~\bibnamefont
  {Dzubiella}}\ and\ \bibinfo {author} {\bibfnamefont {H.}~\bibnamefont
  {L{\"{o}}wen}},\ }\href {\doibase 10.1088/0953-8984/14/40/324} {\bibfield
  {journal} {\bibinfo  {journal} {J.\ Phys.: Condens.\ Matter}\ }\textbf
  {\bibinfo {volume} {14}},\ \bibinfo {pages} {9383} (\bibinfo {year}
  {2002})}\BibitemShut {NoStop}%
\bibitem [{\citenamefont {Chakrabarti}\ \emph {et~al.}(2003)\citenamefont
  {Chakrabarti}, \citenamefont {Dzubiella},\ and\ \citenamefont
  {L{\"{o}}wen}}]{78}%
  \BibitemOpen
  \bibfield  {author} {\bibinfo {author} {\bibfnamefont {J.}~\bibnamefont
  {Chakrabarti}}, \bibinfo {author} {\bibfnamefont {J.}~\bibnamefont
  {Dzubiella}}, \ and\ \bibinfo {author} {\bibfnamefont {H.}~\bibnamefont
  {L{\"{o}}wen}},\ }\href {\doibase 10.1209/epl/i2003-00193-6} {\bibfield
  {journal} {\bibinfo  {journal} {Europhys.\ Lett.}\ }\textbf {\bibinfo
  {volume} {61}},\ \bibinfo {pages} {415} (\bibinfo {year} {2003})}\BibitemShut
  {NoStop}%
\bibitem [{\citenamefont {Chakrabarti}\ \emph {et~al.}(2004)\citenamefont
  {Chakrabarti}, \citenamefont {Dzubiella},\ and\ \citenamefont
  {L{\"{o}}wen}}]{79}%
  \BibitemOpen
  \bibfield  {author} {\bibinfo {author} {\bibfnamefont {J.}~\bibnamefont
  {Chakrabarti}}, \bibinfo {author} {\bibfnamefont {J.}~\bibnamefont
  {Dzubiella}}, \ and\ \bibinfo {author} {\bibfnamefont {H.}~\bibnamefont
  {L{\"{o}}wen}},\ }\href {\doibase 10.1103/PhysRevE.70.012401} {\bibfield
  {journal} {\bibinfo  {journal} {Phys.\ Rev.\ E}\ }\textbf {\bibinfo {volume}
  {70}},\ \bibinfo {pages} {012401} (\bibinfo {year} {2004})}\BibitemShut
  {NoStop}%
\bibitem [{\citenamefont {Leunissen}\ \emph {et~al.}(2005)\citenamefont
  {Leunissen}, \citenamefont {Christova}, \citenamefont {Hynninen},
  \citenamefont {Royall}, \citenamefont {Campbell}, \citenamefont {Imhof},
  \citenamefont {Dijkstra}, \citenamefont {van Roij},\ and\ \citenamefont {van
  Blaaderen}}]{80}%
  \BibitemOpen
  \bibfield  {author} {\bibinfo {author} {\bibfnamefont {M.~E.}\ \bibnamefont
  {Leunissen}}, \bibinfo {author} {\bibfnamefont {C.~G.}\ \bibnamefont
  {Christova}}, \bibinfo {author} {\bibfnamefont {A.-P.}\ \bibnamefont
  {Hynninen}}, \bibinfo {author} {\bibfnamefont {C.~P.}\ \bibnamefont
  {Royall}}, \bibinfo {author} {\bibfnamefont {A.~I.}\ \bibnamefont
  {Campbell}}, \bibinfo {author} {\bibfnamefont {A.}~\bibnamefont {Imhof}},
  \bibinfo {author} {\bibfnamefont {M.}~\bibnamefont {Dijkstra}}, \bibinfo
  {author} {\bibfnamefont {R.}~\bibnamefont {van Roij}}, \ and\ \bibinfo
  {author} {\bibfnamefont {A.}~\bibnamefont {van Blaaderen}},\ }\href {\doibase
  10.1038/nature03946} {\bibfield  {journal} {\bibinfo  {journal} {Nature}\
  }\textbf {\bibinfo {volume} {437}},\ \bibinfo {pages} {235} (\bibinfo {year}
  {2005})}\BibitemShut {NoStop}%
\bibitem [{\citenamefont {Wysocki}\ and\ \citenamefont
  {L{\"{o}}wen}(2004)}]{81}%
  \BibitemOpen
  \bibfield  {author} {\bibinfo {author} {\bibfnamefont {A.}~\bibnamefont
  {Wysocki}}\ and\ \bibinfo {author} {\bibfnamefont {H.}~\bibnamefont
  {L{\"{o}}wen}},\ }\href {\doibase 10.1088/0953-8984/16/41/004} {\bibfield
  {journal} {\bibinfo  {journal} {J.\ Phys.: Condens.\ Matter}\ }\textbf
  {\bibinfo {volume} {16}},\ \bibinfo {pages} {7209} (\bibinfo {year}
  {2004})}\BibitemShut {NoStop}%
\bibitem [{\citenamefont {Wysocki}\ and\ \citenamefont
  {L{\"{o}}wen}(2009)}]{82}%
  \BibitemOpen
  \bibfield  {author} {\bibinfo {author} {\bibfnamefont {A.}~\bibnamefont
  {Wysocki}}\ and\ \bibinfo {author} {\bibfnamefont {H.}~\bibnamefont
  {L{\"{o}}wen}},\ }\href {\doibase 10.1103/PhysRevE.79.041408} {\bibfield
  {journal} {\bibinfo  {journal} {Phys.\ Rev.\ E}\ }\textbf {\bibinfo {volume}
  {79}},\ \bibinfo {pages} {041408} (\bibinfo {year} {2009})}\BibitemShut
  {NoStop}%
\bibitem [{\citenamefont {S{\"{u}}tterlin}\ \emph {et~al.}(2009)\citenamefont
  {S{\"{u}}tterlin}, \citenamefont {Wysocki}, \citenamefont {Ivlev},
  \citenamefont {R{\"{a}}th}, \citenamefont {Thomas}, \citenamefont
  {Rubin-Zuzic}, \citenamefont {Goedheer}, \citenamefont {Fortov},
  \citenamefont {Lipaev}, \citenamefont {Molotkov}, \citenamefont {Petrov},
  \citenamefont {Morfill},\ and\ \citenamefont {L{\"{o}}wen}}]{68}%
  \BibitemOpen
  \bibfield  {author} {\bibinfo {author} {\bibfnamefont {K.~R.}\ \bibnamefont
  {S{\"{u}}tterlin}}, \bibinfo {author} {\bibfnamefont {A.}~\bibnamefont
  {Wysocki}}, \bibinfo {author} {\bibfnamefont {A.~V.}\ \bibnamefont {Ivlev}},
  \bibinfo {author} {\bibfnamefont {C.}~\bibnamefont {R{\"{a}}th}}, \bibinfo
  {author} {\bibfnamefont {H.~M.}\ \bibnamefont {Thomas}}, \bibinfo {author}
  {\bibfnamefont {M.}~\bibnamefont {Rubin-Zuzic}}, \bibinfo {author}
  {\bibfnamefont {W.~J.}\ \bibnamefont {Goedheer}}, \bibinfo {author}
  {\bibfnamefont {V.~E.}\ \bibnamefont {Fortov}}, \bibinfo {author}
  {\bibfnamefont {A.~M.}\ \bibnamefont {Lipaev}}, \bibinfo {author}
  {\bibfnamefont {V.~I.}\ \bibnamefont {Molotkov}}, \bibinfo {author}
  {\bibfnamefont {O.~F.}\ \bibnamefont {Petrov}}, \bibinfo {author}
  {\bibfnamefont {G.~E.}\ \bibnamefont {Morfill}}, \ and\ \bibinfo {author}
  {\bibfnamefont {H.}~\bibnamefont {L{\"{o}}wen}},\ }\href {\doibase
  10.1103/PhysRevLett.102.085003} {\bibfield  {journal} {\bibinfo  {journal}
  {Phys.\ Rev.\ Lett.}\ }\textbf {\bibinfo {volume} {102}},\ \bibinfo {pages}
  {085003} (\bibinfo {year} {2009})}\BibitemShut {NoStop}%
\bibitem [{\citenamefont {Morfill}\ \emph {et~al.}(2012)\citenamefont
  {Morfill}, \citenamefont {Ivlev},\ and\ \citenamefont {Thomas}}]{63}%
  \BibitemOpen
  \bibfield  {author} {\bibinfo {author} {\bibfnamefont {G.~E.}\ \bibnamefont
  {Morfill}}, \bibinfo {author} {\bibfnamefont {A.~V.}\ \bibnamefont {Ivlev}},
  \ and\ \bibinfo {author} {\bibfnamefont {H.~M.}\ \bibnamefont {Thomas}},\
  }\href {\doibase 10.1063/1.4717979} {\bibfield  {journal} {\bibinfo
  {journal} {Phys.\ Plasmas}\ }\textbf {\bibinfo {volume} {19}},\ \bibinfo
  {pages} {055402} (\bibinfo {year} {2012})}\BibitemShut {NoStop}%
\bibitem [{\citenamefont {Zhukhovitskii}\ \emph {et~al.}(2014)\citenamefont
  {Zhukhovitskii}, \citenamefont {Molotkov},\ and\ \citenamefont
  {Fortov}}]{22}%
  \BibitemOpen
  \bibfield  {author} {\bibinfo {author} {\bibfnamefont {D.~I.}\ \bibnamefont
  {Zhukhovitskii}}, \bibinfo {author} {\bibfnamefont {V.~I.}\ \bibnamefont
  {Molotkov}}, \ and\ \bibinfo {author} {\bibfnamefont {V.~E.}\ \bibnamefont
  {Fortov}},\ }\href {\doibase 10.1063/1.4881473} {\bibfield  {journal}
  {\bibinfo  {journal} {Phys.\ Plasmas}\ }\textbf {\bibinfo {volume} {21}},\
  \bibinfo {pages} {063701} (\bibinfo {year} {2014})}\BibitemShut {NoStop}%
\bibitem [{\citenamefont {Zhukhovitskii}(2015)}]{64}%
  \BibitemOpen
  \bibfield  {author} {\bibinfo {author} {\bibfnamefont {D.~I.}\ \bibnamefont
  {Zhukhovitskii}},\ }\href {\doibase 10.1103/PhysRevE.92.023108} {\bibfield
  {journal} {\bibinfo  {journal} {Phys.\ Rev.\ E}\ }\textbf {\bibinfo {volume}
  {92}},\ \bibinfo {pages} {023108} (\bibinfo {year} {2015})}\BibitemShut
  {NoStop}%
\bibitem [{\citenamefont {Mott-Smith}\ and\ \citenamefont
  {Langmuir}(1926)}]{54}%
  \BibitemOpen
  \bibfield  {author} {\bibinfo {author} {\bibfnamefont {H.~M.}\ \bibnamefont
  {Mott-Smith}}\ and\ \bibinfo {author} {\bibfnamefont {I.}~\bibnamefont
  {Langmuir}},\ }\href {\doibase 10.1103/PhysRev.28.727} {\bibfield  {journal}
  {\bibinfo  {journal} {Phys.\ Rev.}\ }\textbf {\bibinfo {volume} {28}},\
  \bibinfo {pages} {727} (\bibinfo {year} {1926})}\BibitemShut {NoStop}%
\bibitem [{\citenamefont {Allen}(1992)}]{55}%
  \BibitemOpen
  \bibfield  {author} {\bibinfo {author} {\bibfnamefont {J.~E.}\ \bibnamefont
  {Allen}},\ }\href {\doibase 10.1088/0031-8949/45/5/013} {\bibfield  {journal}
  {\bibinfo  {journal} {Phys.\ Scr.}\ }\textbf {\bibinfo {volume} {45}},\
  \bibinfo {pages} {497} (\bibinfo {year} {1992})}\BibitemShut {NoStop}%
\bibitem [{\citenamefont {Zhukhovitskii}\ \emph {et~al.}(2012)\citenamefont
  {Zhukhovitskii}, \citenamefont {Fortov}, \citenamefont {Molotkov},
  \citenamefont {Lipaev}, \citenamefont {Naumkin}, \citenamefont {Thomas},
  \citenamefont {Ivlev}, \citenamefont {Schwabe},\ and\ \citenamefont
  {Morfill}}]{49}%
  \BibitemOpen
  \bibfield  {author} {\bibinfo {author} {\bibfnamefont {D.~I.}\ \bibnamefont
  {Zhukhovitskii}}, \bibinfo {author} {\bibfnamefont {V.~E.}\ \bibnamefont
  {Fortov}}, \bibinfo {author} {\bibfnamefont {V.~I.}\ \bibnamefont
  {Molotkov}}, \bibinfo {author} {\bibfnamefont {A.~M.}\ \bibnamefont
  {Lipaev}}, \bibinfo {author} {\bibfnamefont {V.~N.}\ \bibnamefont {Naumkin}},
  \bibinfo {author} {\bibfnamefont {H.~M.}\ \bibnamefont {Thomas}}, \bibinfo
  {author} {\bibfnamefont {A.~V.}\ \bibnamefont {Ivlev}}, \bibinfo {author}
  {\bibfnamefont {M.}~\bibnamefont {Schwabe}}, \ and\ \bibinfo {author}
  {\bibfnamefont {G.~E.}\ \bibnamefont {Morfill}},\ }\href {\doibase
  10.1103/PhysRevE.86.016401} {\bibfield  {journal} {\bibinfo  {journal}
  {Phys.\ Rev.\ E}\ }\textbf {\bibinfo {volume} {86}},\ \bibinfo {pages}
  {016401} (\bibinfo {year} {2012})}\BibitemShut {NoStop}%
\bibitem [{\citenamefont {Epstein}(1924)}]{33}%
  \BibitemOpen
  \bibfield  {author} {\bibinfo {author} {\bibfnamefont {P.}~\bibnamefont
  {Epstein}},\ }\href {\doibase 10.1103/PhysRev.23.710} {\bibfield  {journal}
  {\bibinfo  {journal} {Phys.\ Rev.}\ }\textbf {\bibinfo {volume} {23}},\
  \bibinfo {pages} {710} (\bibinfo {year} {1924})}\BibitemShut {NoStop}%
\bibitem [{\citenamefont {Ivlev}\ and\ \citenamefont
  {Zhukhovitskii}(2012)}]{52}%
  \BibitemOpen
  \bibfield  {author} {\bibinfo {author} {\bibfnamefont {A.~V.}\ \bibnamefont
  {Ivlev}}\ and\ \bibinfo {author} {\bibfnamefont {D.~I.}\ \bibnamefont
  {Zhukhovitskii}},\ }\href {\doibase 10.1063/1.4750070} {\bibfield  {journal}
  {\bibinfo  {journal} {Phys.\ Plasmas}\ }\textbf {\bibinfo {volume} {19}},\
  \bibinfo {pages} {093703} (\bibinfo {year} {2012})}\BibitemShut {NoStop}%
\bibitem [{\citenamefont {Zhukhovitskii}\ \emph {et~al.}(2013)\citenamefont
  {Zhukhovitskii}, \citenamefont {Ivlev}, \citenamefont {Fortov},\ and\
  \citenamefont {Morfill}}]{53}%
  \BibitemOpen
  \bibfield  {author} {\bibinfo {author} {\bibfnamefont {D.~I.}\ \bibnamefont
  {Zhukhovitskii}}, \bibinfo {author} {\bibfnamefont {A.~V.}\ \bibnamefont
  {Ivlev}}, \bibinfo {author} {\bibfnamefont {V.~E.}\ \bibnamefont {Fortov}}, \
  and\ \bibinfo {author} {\bibfnamefont {G.~E.}\ \bibnamefont {Morfill}},\
  }\href {\doibase 10.1103/PhysRevE.87.063108} {\bibfield  {journal} {\bibinfo
  {journal} {Phys.\ Rev.\ E}\ }\textbf {\bibinfo {volume} {87}},\ \bibinfo
  {pages} {063108} (\bibinfo {year} {2013})}\BibitemShut {NoStop}%
\bibitem [{\citenamefont {Naumkin}\ \emph {et~al.}(2016)\citenamefont
  {Naumkin}, \citenamefont {Zhukhovitskii}, \citenamefont {Molotkov},
  \citenamefont {Lipaev}, \citenamefont {Fortov}, \citenamefont {Thomas},
  \citenamefont {Huber},\ and\ \citenamefont {Morfill}}]{69}%
  \BibitemOpen
  \bibfield  {author} {\bibinfo {author} {\bibfnamefont {V.~N.}\ \bibnamefont
  {Naumkin}}, \bibinfo {author} {\bibfnamefont {D.~I.}\ \bibnamefont
  {Zhukhovitskii}}, \bibinfo {author} {\bibfnamefont {V.~I.}\ \bibnamefont
  {Molotkov}}, \bibinfo {author} {\bibfnamefont {A.~M.}\ \bibnamefont
  {Lipaev}}, \bibinfo {author} {\bibfnamefont {V.~E.}\ \bibnamefont {Fortov}},
  \bibinfo {author} {\bibfnamefont {H.~M.}\ \bibnamefont {Thomas}}, \bibinfo
  {author} {\bibfnamefont {P.}~\bibnamefont {Huber}}, \ and\ \bibinfo {author}
  {\bibfnamefont {G.~E.}\ \bibnamefont {Morfill}},\ }\href {\doibase
  10.1103/PhysRevE.94.033204} {\bibfield  {journal} {\bibinfo  {journal}
  {Phys.\ Rev.\ E}\ }\textbf {\bibinfo {volume} {94}},\ \bibinfo {pages}
  {033204} (\bibinfo {year} {2016})}\BibitemShut {NoStop}%
\bibitem [{\citenamefont {Khrapak}\ \emph {et~al.}(2005)\citenamefont
  {Khrapak}, \citenamefont {Ratynskaia}, \citenamefont {Zobnin}, \citenamefont
  {Usachev}, \citenamefont {Yaroshenko}, \citenamefont {Thoma}, \citenamefont
  {Kretschmer}, \citenamefont {H{\"{o}}fner}, \citenamefont {Morfill},
  \citenamefont {Petrov},\ and\ \citenamefont {Fortov}}]{60}%
  \BibitemOpen
  \bibfield  {author} {\bibinfo {author} {\bibfnamefont {S.~A.}\ \bibnamefont
  {Khrapak}}, \bibinfo {author} {\bibfnamefont {S.~V.}\ \bibnamefont
  {Ratynskaia}}, \bibinfo {author} {\bibfnamefont {A.~V.}\ \bibnamefont
  {Zobnin}}, \bibinfo {author} {\bibfnamefont {A.~D.}\ \bibnamefont {Usachev}},
  \bibinfo {author} {\bibfnamefont {V.~V.}\ \bibnamefont {Yaroshenko}},
  \bibinfo {author} {\bibfnamefont {M.~H.}\ \bibnamefont {Thoma}}, \bibinfo
  {author} {\bibfnamefont {M.}~\bibnamefont {Kretschmer}}, \bibinfo {author}
  {\bibfnamefont {H.}~\bibnamefont {H{\"{o}}fner}}, \bibinfo {author}
  {\bibfnamefont {G.~E.}\ \bibnamefont {Morfill}}, \bibinfo {author}
  {\bibfnamefont {O.~F.}\ \bibnamefont {Petrov}}, \ and\ \bibinfo {author}
  {\bibfnamefont {V.~E.}\ \bibnamefont {Fortov}},\ }\href {\doibase
  10.1103/PhysRevE.72.016406} {\bibfield  {journal} {\bibinfo  {journal}
  {Phys.\ Rev.\ E}\ }\textbf {\bibinfo {volume} {72}},\ \bibinfo {pages}
  {016406} (\bibinfo {year} {2005})}\BibitemShut {NoStop}%
\bibitem [{\citenamefont {Khrapak}\ \emph {et~al.}(2012)\citenamefont
  {Khrapak}, \citenamefont {Klumov}, \citenamefont {Huber}, \citenamefont
  {Molotkov}, \citenamefont {Lipaev}, \citenamefont {Naumkin}, \citenamefont
  {Ivlev}, \citenamefont {Thomas}, \citenamefont {Schwabe}, \citenamefont
  {Morfill}, \citenamefont {Petrov}, \citenamefont {Fortov}, \citenamefont
  {Malentschenko},\ and\ \citenamefont {Volkov}}]{43}%
  \BibitemOpen
  \bibfield  {author} {\bibinfo {author} {\bibfnamefont {S.~A.}\ \bibnamefont
  {Khrapak}}, \bibinfo {author} {\bibfnamefont {B.~A.}\ \bibnamefont {Klumov}},
  \bibinfo {author} {\bibfnamefont {P.}~\bibnamefont {Huber}}, \bibinfo
  {author} {\bibfnamefont {V.~I.}\ \bibnamefont {Molotkov}}, \bibinfo {author}
  {\bibfnamefont {A.~M.}\ \bibnamefont {Lipaev}}, \bibinfo {author}
  {\bibfnamefont {V.~N.}\ \bibnamefont {Naumkin}}, \bibinfo {author}
  {\bibfnamefont {A.~V.}\ \bibnamefont {Ivlev}}, \bibinfo {author}
  {\bibfnamefont {H.~M.}\ \bibnamefont {Thomas}}, \bibinfo {author}
  {\bibfnamefont {M.}~\bibnamefont {Schwabe}}, \bibinfo {author} {\bibfnamefont
  {G.~E.}\ \bibnamefont {Morfill}}, \bibinfo {author} {\bibfnamefont {O.~F.}\
  \bibnamefont {Petrov}}, \bibinfo {author} {\bibfnamefont {V.~E.}\
  \bibnamefont {Fortov}}, \bibinfo {author} {\bibfnamefont {Y.}~\bibnamefont
  {Malentschenko}}, \ and\ \bibinfo {author} {\bibfnamefont {S.}~\bibnamefont
  {Volkov}},\ }\href {\doibase 10.1103/PhysRevE.85.066407} {\bibfield
  {journal} {\bibinfo  {journal} {Phys.\ Rev.\ E}\ }\textbf {\bibinfo {volume}
  {85}},\ \bibinfo {pages} {066407} (\bibinfo {year} {2012})}\BibitemShut
  {NoStop}%
\end{thebibliography}
\end{document}